\documentclass[superscript address,
reprint,
 amsmath,amssymb,
 aps,
]{revtex4-2}

\usepackage{graphicx} 
\usepackage{graphics}
\usepackage{epstopdf}
\usepackage{dcolumn}
\usepackage{bm}
\usepackage{amsmath}
\usepackage{breqn}
\DeclareMathOperator{\Tr}{Tr}
\usepackage{amssymb}
\usepackage{mathrsfs}
\usepackage{amsfonts}
\usepackage{physics}
\usepackage{mathtools}
\usepackage{subfig}
\makeatletter
\def\cat@comma@active{\catcode`\,12}%
\makeatother

\usepackage[utf8]{inputenc}
\usepackage[english]{babel}
\usepackage[T1]{fontenc}
\usepackage{microtype}

\usepackage[rightcaption]{sidecap}

\usepackage{lmodern}
\usepackage{amsmath}
\usepackage{amssymb}
\usepackage{mathrsfs}
\usepackage{amsfonts}
\usepackage{physics}
\usepackage{float}

\usepackage{times}
\usepackage{xcolor}
\usepackage{subfig,graphicx}

\usepackage[colorinlistoftodos]{todonotes}

\begin{document}
\preprint{APS/123-QED}
\title{Pairwise quantum correlations in four-level quantum dot systems}

\author{Sanaa ABAACH}
\email{sanaa_abaach@um5.ac.ma}
\author{Morad EL BAZ}
\email{morad.elbaz@um5.ac.ma}
\affiliation{ESMaR, Faculty of Sciences, Mohammed V University in Rabat, Morocco.}
\author{Mustapha Faqir}
\email{mustapha.faqir@uir.ac.ma}
\affiliation{Université internationale de Rabat, Aerospace engineering school, LERMA lab, Morocco.}


\begin{abstract}
In this paper we assume quantum dots can be assimilated to Fermi Hubbard sites when the Coulomb interaction between electrons is higher compared to their tunneling. The study of pairwise entanglement in a small size array of quantum dots allows to model each pair as a quadrit-quadrit system (4 $\times$ 4 mixed state) instead of the more common and simplistic approach of describing it in quantum information as a qubit-qubit system. We study the effect of Coulomb interaction and temperature on pairwise entanglement as well as on quantum coherence and total correlations. The crucial results of this study are that entanglement resists better the increase in temperature when the Coulomb interaction is stronger. Moreover, we successfully explain the behavior of these correlations in terms of the energy spectrum, namely the ground state degeneracy and the state energy difference.
\end{abstract}

\maketitle


\section{Introduction}

Quantum entanglement has been a major part of various debates and discussions from the earliest days of quantum mechanics \cite{einstein,bell}. It has provided an immense impact in various disciplines e.g. quantum optics \cite{optics}, condensed matter \cite{CM}, quantum information \cite{QI} etc. The study of entanglement in condensed matter model systems has taken advantage of the rapid advancement in quantum information, where the many-body entangled states became a powerful and active subject of research, especially after discovering that entanglement could serve as a new revolution for quantum critical phenomena \cite{PT1,PT2}.

Condensed matter is at the heart of nano-fabrication, nano-technologies and mesoscopic physics in general. One of the potent electronic phenomena studied in this field is the quantum confinement effect of electrons which is fundamental for the fabrication of quantum dot systems \cite{QDF}. These man-made nano-crystal materials have the quantum properties of single atoms due to their discrete energy levels. After the discovery that quantum dots can robustly serve quantum information processing especially in building quantum computers \cite{CC}, many researches were dedicated to study the properties of such nanostructures. Despite all this, modeling these latter is still a hard task in computation in general. Consequently lattice models, such as the commonly known Hubbard model, are frequently used as an approximation \cite{teleportation,Approx1,Approx2}. Because of the prevailing electron-electron interactions observed in quantum dot systems (yielding to the well known Coulomb blockade phenomenon) the Hubbard model is considered a suitable candidate for describing quantum dot systems. Recently, a scalable set of experimental tools were realized using an array of three semiconductor quantum dots and have validated the simulation of Fermi–Hubbard physics using semiconductor quantum dots \cite{simulation}. Furthermore it has been shown that the entanglement shared between the quantum dots can be modeled by the one-dimensional Hubbard model and it was found suitable for the calculation of single-site entanglement \cite{Approx}.

Quantum simulator is a particular purpose device that is designed to study and simulate quantum many-body problems that are impossible to solve on a classical computer . Several experiments and theoretical studies have demonstrated that such a device is possible to build for the Fermi-Hubbard model using quantum dot array systems \cite{simulation, Manousakis2002AQA, PhysRevB.78.075320, doi:10.1002/andp.201300124}.

Moreover, it has been widely shown that quantum dot systems are one of the most tunable and efficient electrically controlled quantum systems \cite{RevModPhys.75.1} . This tunability of their design allows them to be a good platform that naturally obeys the Fermionic Hubbard physics in strong coupling interaction regime at low temperature (when cooled down to dilution temperatures). Typical parameter values  of coupling energy ($u \approx 1 \; meV$), inter-dot tunnel coupling ($t \approx 10 - 100 \; \mu eV$) and dilution refrigerator temperature ($k_BT \approx 1 - 10\;  \mu eV$) provide accessibility to one of the most interesting regimes of the Hubbard Hamiltonian \cite{RevModPhys.75.1, RevModPhys.79.1217}. From an experimental point of view, the most commonly studied and experimentally employed quantum dot devices are usually on GaAs/AlGaAs semiconductor heterostructures that grow by molecular-beam epitaxy \cite{simulation, Manousakis2002AQA, PhysRevB.78.075320, doi:10.1002/andp.201300124, RevModPhys.75.1}. Free electrons are strongly confined (in one direction) to the interface between GaAs and AlGaAs, producing thus a two-dimensional electron gas (2DEG). The confinement in the remaining two dimensions is performed by locally depleting the 2DEG, via metal gate electrodes on the surface of the heterostructure. The control of Fermi–Hubbard parameters is reached by adjusting the potential landscape in the 2DEG using the gate electrodes. These gates include plunger gates and barrier gates that are designed to tune the single-particle energy offsets, the chemical potential of individual dots, the tunnel couplings ($t$) between two dots as well as the on-site ($u$) and the inter-site Coulomb interaction energies. Another excellent technological platform increasingly used during the last decade is Silicon-based quantum devices. Because of the low nuclear spin density as well as the weak spin orbit coupling in Silicon, this latter represents an ideal environment for spins in the solid state. Si MOSFETs (Metal-Oxide-Semiconductor Field Effect Transistors) and Si/SiGe heterostructures are among the accurate quantum dot devices in experiment, that provide longer spin relaxation and coherence times contrary to GaAs/AlGaAs heterostructures \cite{Silicon}. Silicon quantum dot systems are quantitatively described by the Hubbard model approach \cite{siliconHubbard} and the same mechanism of electron confinement, previously described for GaAs/AlGaAs, is generally adopted for Silicon-based heterostructure quantum dots.

The Hubbard model was introduced in the 1960s as a simple approximation to describe the motion of interacting spin-1/2 fermions in a lattice.  Recently, it has been found that this model can be a good candidate for universal quantum computation and has been explored to construct a universal set of quantum gates whose purpose is the construction of a quantum computer based on interacting fermions \cite{CC1}. A crucial point characterizing the Hubbard model is its ground state that represents a natural source of entanglement and hence it can be a significant resource for quantum information processing such as quantum teleportation. It has been successfully demonstrated that at half filling, in the metallic state, the ground state describing two sites of the Hubbard model is maximally entangled and provides a good quantum channel to teleport a qubit between two parties \cite{teleportation}. In view of the fact that the quantum states of such a model are classified in the category of qudits, a proper description of the whole system represents a big challenge for quantum information to find a suitable quantification of entanglement for such multipartite quantum systems, particularly in the case of mixed states. Although, 
entanglement of formation and concurrence are well-defined measures of entanglement for bipartite systems \cite{wootters}, with the growth of the dimension of the subsystems, they become exceedingly hard in computation. For multipatite systems in higher dimensions the quantification of entanglement is still an open question in quantum information, but particular lower bounds of concurrence have been suggested recently providing an estimation of this multipartite entanglement \cite{lower}.

Finite-size systems have attracted a great deal of attention in the last decades because of the focus on nanotechnological applications, besides they are appropriate for numerical treatment that can be extrapolated to infinite lattices. The main goal of this paper is to quantify and study the behavior of entanglement and quantum coherence between various pairs, located in a small size array of quantum dots under the effect of Coulomb interaction and temperature, basing on the one dimensional Fermi-Hubbard model. In the next section we introduce the model and the principal measures of entanglement for the Hubbard chain. We  also define the lower bound of concurrence for the mixed state describing a pair of quantum dots. In addition to that we introduce quantum coherence as well as total correlations. In the third section we present and discuss our results about the effect of temperature and the Coulomb coupling on the pair correlations. We start by analyzing the ground state entanglement and how this latter can be influenced by the degeneracy of the states, then we study the pair correlations at zero temperature, finite temperature and finally high temperature. We give an adequate explanation for the  behavior of these correlations at finite and high temperature based on the state energy difference in the energy spectrum. In the last section, a conclusion and core results of this paper are summarized.

\section{Model and Formalism}
\subsection{Fermi-Hubbard Model}
\quad A chain of quantum dots, as artificial atoms, can be theoretically described by means of the one-dimensional Fermi-Hubbard model. If we Assume that the hopping is bounded by the nearest-neighbor lattice sites, the simplest expression of the Hamiltonian corresponding to the model is formulated as follows
\begin{equation}
H = -t \sum_{i,\sigma} \left( c_{i,\sigma}^{\dag} c_{i+1,\sigma}+ c_{i+1,\sigma}^{\dag} c_{i,\sigma} \right) + u \sum_{i} n_ {i,\uparrow} n_{i,\downarrow}.    
\label{ham}
\end{equation}
The first term in the Hamiltonian represents the Kinetic energy that takes account of the tunneling between neighboring sites, whereby each electron at site $i$ with a given spin $\sigma= \{ \uparrow,\downarrow \} $ can leave its position in order to occupy the nearest neighboring site $i+1$ (and \textit{vice versa}), with $t$ being the hopping amplitude. The second term represents the on-site electron-electron Coulomb interaction $u$. Moreover,
we assume that only one energy level is allowed to electrons in each quantum dot which means that only the $s$-orbital is taken into account therefore each quantum dot is able to hold, up to two electrons with opposite spins according to the Pauli exclusion principle.

\begin{figure}
  \centering
  \subfloat[]{\includegraphics[width=.48\linewidth]{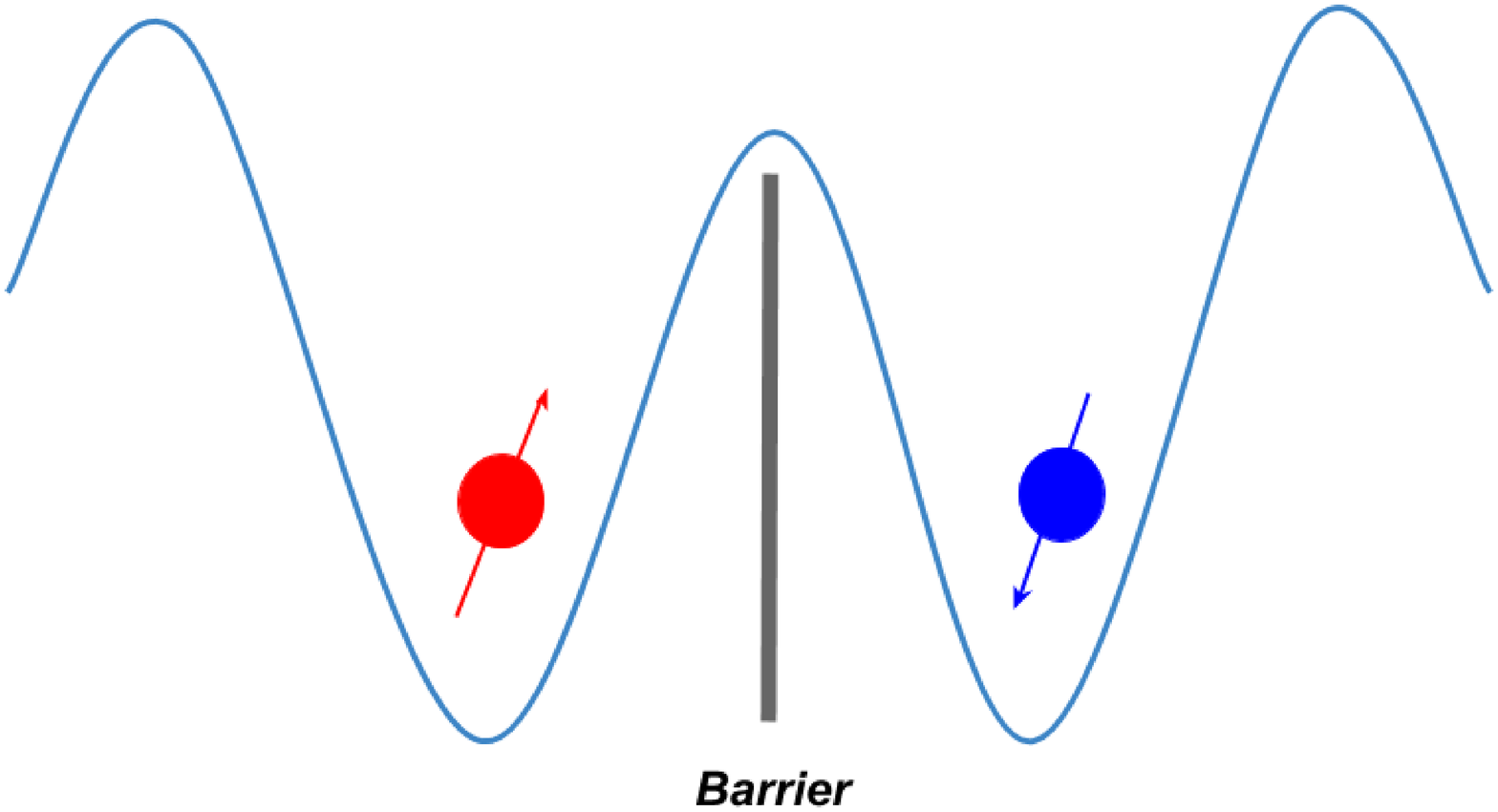}
  \label{fig:img0a}} 
  \subfloat[]{\includegraphics[width=.48\linewidth]{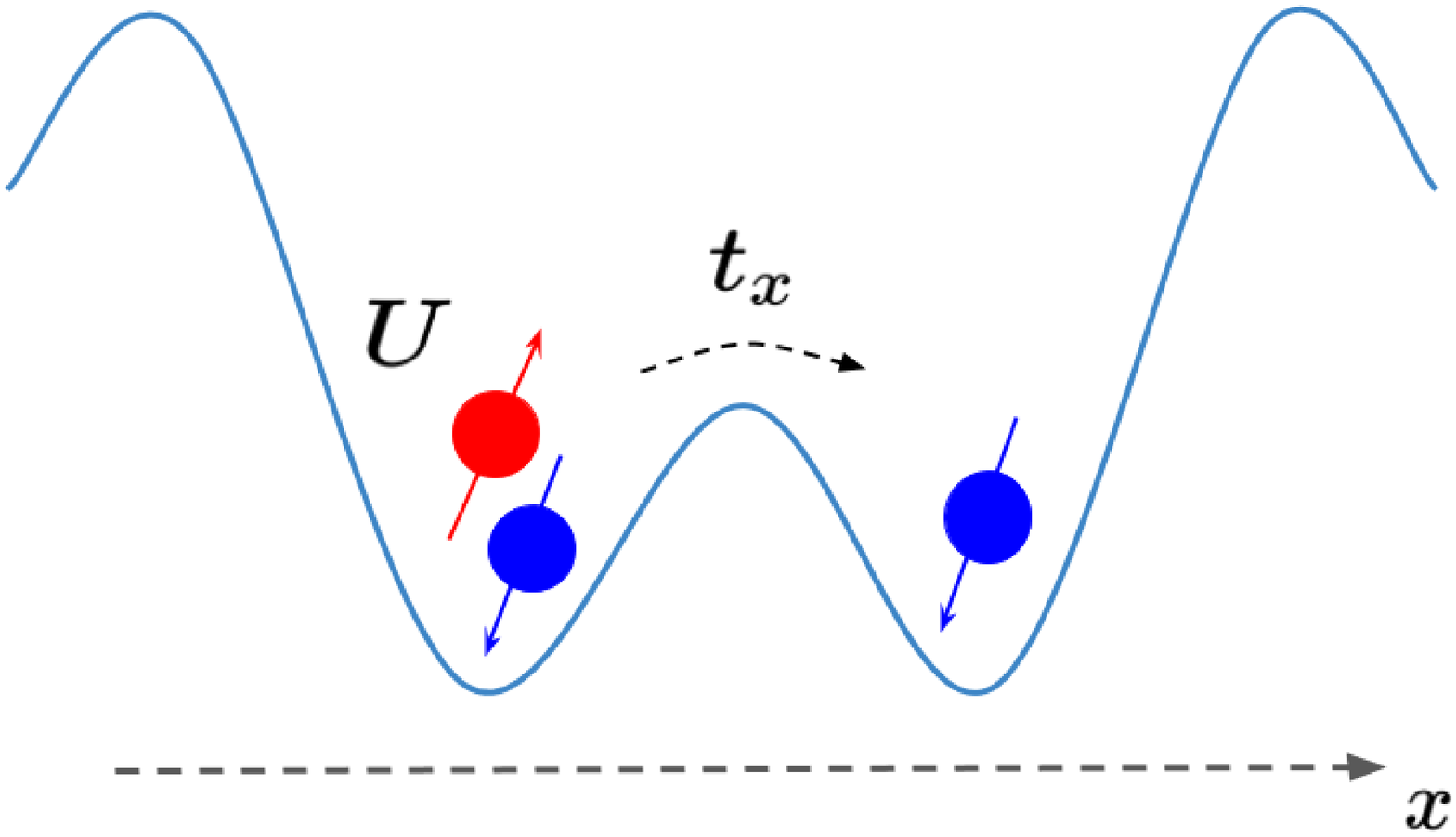}
  \label{fig:img0b}}
  \caption{ \it Representative scheme showing the confinement of electrons in the Fermi-Hubbard sites when a) the tunneling is blocked and b) when it is unblocked (the electrons are free to tunnel between the site in the $x$ direction).   }
  \label{fig : img0}
\end{figure}

The Fermi Hubbard sites can be assimilated to quantum dots when $u/t$ takes very large values \cite{simulation}. In fact, when the repulsion interaction $u$ within sites is very strong and the tunneling of electrons between the sites is blocked this is analogous to the confinement effect in the Fermi-Hubbard model. The situation is similar to the creation of potential barriers between the sites which prevent electrons to move outside (see Figure (\ref{fig:img0a}). Experimentally such barriers can be produced by modulating the potentials using gate electrodes in order to control the tunneling of electrons between quantum dots \cite{simulation}. For weak coupling $u/t$, \textit{i.e.} strong tunneling, the system can be assimilated to coupled quantum wires where the confinement of electrons is reached in the $y-z$ plane while they can freely tunnel through the $x$ direction (Figure (\ref{fig:img0b})). Experimentally the inter-dot Coulomb interaction (generally denoted $v$) is taken into account in quantum dot systems but has to be as small as possible ($t<v<u$) \cite{simulation}. It has been shown that even in the well isolated quantum dots case ($t=0$), a significant inter-dot Coulomb interaction could destroy the Coulomb blockade \cite{Blocked} that is a fundamental phenomenon observed while creating quantum dots, causing thus a current fluctuations (electrons transport). Because of the reasoning mentioned above our theoretical study is limited to consider the case where the inter-dot Coulomb interaction is neglected. However, if this interaction is to be taken into account, one has to use the extended Hubbard model instead \cite{simulation}.

\subsection{Entanglement measures in Hubbard chain\label{mesure}}
Entanglement is a powerful quantum phenomenon that allows two or more systems to be linked in such a way that any action on one subsystem influences the other instantaneously. Moreover, the Hubbard model is characterized by an entangled ground state. In a quantum dot, electrons have four possibilities in occupying a single site: $(\ket{0},\ket{\uparrow},\ket{\downarrow},\ket{\uparrow \downarrow})$. As a result an array of $N$ quantum dots can be defined by the ground state $\ket{\psi}\in \mathcal{H}$, where $\mathcal{H}$ is the Hilbert space of dimension $4^{N}$. The quantification of entanglement in such arrays has been limited, so far \cite{singlesite,block}, to computing the amount of entanglement in pure states, by means of the Van Neumann entropy $E_{k}= -\Tr
(\rho_{k}\log_{2}\rho_{k})$, where  $ \rho_{k}= \Tr
_{N-k}{(\ket{\psi}\bra{\psi})}$ is the state of the subsystem composed of $ k=\{1,2,...,N/2\}$ quantum dots  and $\Tr
_{N-k}$ means the partial trace over the remaining $N-k$ sites. When $k=1$ this gives the so called local entanglement between a single site and the remaining $N-1$ sites. In this case, $\rho_{k}$ is given by \cite{singlesite} 
\begin{equation}
    \rho_{k}=\omega_{0} \ket{0} \bra{0}+\omega_{\uparrow} \ket{\uparrow} \bra{\uparrow}+\omega_{\downarrow} \ket{\downarrow}\bra{\downarrow} +\omega_{\uparrow \downarrow}\ket{\uparrow \downarrow}\bra{\uparrow \downarrow},
    \label{gho}
\end{equation}
where 
\begin{equation}
\begin{aligned}
    \omega_{\uparrow \downarrow} &=    \Tr(n_{k,\uparrow}n_{k,\downarrow} \rho_{k}) = < n_{k,\uparrow}n_{k,\downarrow} >, \\
     \omega_{\uparrow} &= < n_{k,\uparrow}> -  \omega_{\uparrow \downarrow}, \\
     \omega_{\downarrow} &= <n_{k,\downarrow}> -  \omega_{\uparrow \downarrow}, \\
     \omega_{0} &= 1-( \omega_{\uparrow \downarrow} +\omega_{\uparrow} + \omega_{\downarrow}). 
\end{aligned}   
\label{omega}
\end{equation}
The $U(1)$ and $SU(2)$ symmetries in the Hubbard model means that the off-diagonal elements of Eq.~\eqref{gho} are equal to zero.
The four diagonal elements defined in Eq.~\eqref{omega}, thus play an essential role in finding the physical and quantum properties of the chain. Accordingly, the local entanglement,
\begin{equation}
    E_{k}= - \omega_{0}\log_{2}\omega_{0}- \omega_{\uparrow}\log_{2}\omega_{\uparrow}- \omega_{\downarrow}\log_{2}\omega_{\downarrow} - \omega_{\uparrow \downarrow}\log_{2}\omega_{\uparrow \downarrow},
\end{equation}
depends solely on these four quantities and it presents a key role in the understanding of the system \cite{PT2}.

When $k\ge2$, the von Neumann entropy defined above yields the Block-Block entanglement between $k$ sites and the remaining $N-k$ sites of the chain. It has been demonstrated that this kind of entanglement gives more information about the system because it contains nonlocal correlations compared to the local single site entanglement \cite{block}.

Another way of measuring quantum entanglement for such composite systems can be achieved by employing the generalized concurrence defined for qudits \cite{purestate}, which for a given pure state $\ket{\psi}\in \mathcal{H}_{1}\otimes \mathcal{H}_{2} \otimes...\otimes\mathcal{H}_{N}$ where $\mathcal{H}_{i}$ is the Hilbert space of $d_{i}$ dimensions, has the following form :
\begin{equation}
    C_{N}(\ket{\psi} \bra{\psi})=2^{1-\frac{N}{2}}\sqrt{(2^N -2)-\sum_{\alpha}\Tr(\rho_{\alpha}^2)}
    \label{concr},
\end{equation} where $\alpha$ labels all the possible $2^{N}-2$ subsets of the $N$ particle system. 

As a matter of fact Eq.~\eqref{concr} combines the two types of entanglement mentioned above ($k=1$ and $k\ge 2$) because it takes account of all the possible partitions of the system. Consequently it may enrich the information about the quantum and physical properties of the system. The down side of the previous measures of entanglement, is their restricted validity for pure states only.

\subsection{Lower bound of concurrence}

Acquaintance about the amount of entanglement shared between a pair, triple or generally multiple sites (instead of blocks of sites as discussed in the previous subsection) is also of valuable interest. However, an appropriate measure of multipartite entanglement for mixed states in higher dimensions ($d\ge3$, \textit{i.e. qudits}), is not well defined yet \cite{REVmxd}. As a way to circumvent this difficulty, one can calculate the lower bound of concurrence for such systems  \cite{lower}. Despite the fact that this approach does not give the complete knowledge about the quantity of entanglement, it nevertheless gives valuable information that is not obtainable otherwise. The robustness of this lower bound of concurrence is manifested in the fact that it can detect mixed entangled states with a positive partial transpose \cite{CLBPPT} and that for fully separable multipartite state it is equal to zero . Based on the general definition of the lower bound of concurrence one derives simpler expressions for bipartite state that will be of interest for us in this paper as we are interested in the study of pairwise entanglement.  

A pair of two sites is defined by the mixed state $ \rho_{i,j}$ ($i<j$) such that $\rho_{i,j}=\Tr_{N-2}{(\ket{\psi}\bra{\psi})}$, with $\Tr_{N-2}$ is tracing over all sites except the $i^{\text{\tiny th}}$ and $j^{\text{\tiny th}}$ sites.
For an arbitrary $d\times d$ dimension the concurrence  $C(\rho_{i,j})$ satisfies \cite{lower}
\begin{equation}
  \tau_{2}( \rho_{i,j})=\frac{d}{2(d-1)}\sum_{\alpha}^{\frac{d(d-1)}{2}}\sum_{\beta}^{\frac{d(d-1)}{2}} C_{\alpha\beta}^{2}\le C^{2}(\rho_{i,j}),
  \label{pairlow}
\end{equation}
where
\begin{equation}
C_{\alpha\beta}=max\{0,\lambda_{\alpha\beta}^{(1)}-\lambda_{\alpha\beta}^{(2)}-\lambda_{\alpha\beta}^{(3)}-\lambda_{\alpha\beta}^{(4)} \}. \\
\end{equation}
The $ \lambda_{\alpha\beta}^{(\hbox{a})} $ are the square roots of the non-zero eigenvalues of the non-Hermitian matrix $ \rho_{i,j}\tilde{\rho}_{(i,j)\alpha\beta} $ such that 
$ \lambda_{\alpha\beta}^{(\hbox{a})}> \lambda_{\alpha\beta}^{(\hbox{a}+1)} $ for $ 1\le \hbox{a}\le 3$ and
\begin{equation}
   \tilde{\rho}_{(i,j)\alpha\beta}=(G_{\alpha}\otimes G_{\beta})\rho_{i,j}^{*}(G_{\alpha}\otimes G_{\beta}),
\end{equation}
with $G_{\alpha}$ being the  $\alpha^{\text{\tiny th}}$ element of the group $SO(d)$ constructed by $\frac{d(d-1)}{2}$ generators. $G_{\beta}$ is defined similarly since the two subsystems have the same dimension $d$.

According to Eq.~\eqref{pairlow} the lower bound of concurrence is  $\sqrt{\tau_{2}( \rho_{i,j})}$ . When $\tau_{2}( \rho_{i,j})=0$ the state is separable but when $\tau_{2}( \rho_{i,j})>0$ this indicates that inside the quantum state some kinds of entanglement are detected. For multiple sites the task will be hard in computation with the growth of the dimension especially that in our case each subsystem of the Hubbard model has a dimension $d=4$. Therefore if one has at least three subsystems for example \cite{lower}, the number of generators constructing the group $SO(d^{2})\equiv SO(16)$ is $120$ which is a very large number for computation.

\subsection{Quantum coherence and total correlations}

One of the most important phenomena and pillars of quantum mechanics is the superposition of quantum states. This phenomenon is even more fundamental than entanglement as the latter is a direct consequence of the former when applied to composite systems. Accordingly, the evaluation and quantification of such property is an essential task of quantum information theory as it give more general and fundamental information than entanglement. A rigorous measure of quantum coherence is defined as the distance produced by the relative entropy between a given quantum state with its nearest incoherent one. It is demonstrated that the nearest incoherent state is nothing else than the corresponding diagonal matrix for which all the  off-diagonal elements are equal to zero. A simple analytical expression of quantum coherence has the form \cite{coher}
\begin{equation}
    C(\rho)=E(\rho_{diag})-E(\rho),
\label{coherence}    
\end{equation}
where $\rho_{diag}$ is the diagonal matrix of the density matrix $\rho$ and $E(\rho)$ is the von Neumann entropy defined in subsection \ref{mesure}. 

Alternatively, based on the von Neumann entropy, one can define the quantum mutual information:
\begin{equation}
    I(\rho_{AB}) = E(\rho_{A})+ E(\rho_{B})- E(\rho_{AB}),
\end{equation}
which is widely adopted as a quantifier for the total (quantum and classical included) correlations, in a bipartite quantum state $\rho_{AB}$.  For maximally entangled pure state $\rho_{AB}$ the Van Neuman entropy satisfies $E(\rho_{A}) = E(\rho_{B}) = \log_{2}(d)$, where $d$ is the dimension of the subsystems $A$ and $B$. This implies that as the dimension increases, the maximum of entanglement increases and consequently quantum coherence (Eq.~\eqref{coherence}) achieves its maximum $\log_{2}(d)$ (because $E(\rho_{AB}) = 0$) while the mutual information reaches $ 2\log_{2}(d)$. Since in our case we consider the quantum dots as 4-dimensional systems the maximum attainable value for coherence is 2 and for the mutual information it is 4. However, it is worth mentioning that these maxima can be surpassed if considering bipartitions beyond 4x4 dimensions. On the other hand, for totally mixed states (which are separable), total correlations including coherence and entanglement vanish.

\section{Results and discussion}

\begin{figure}[ht]\centering
\subfloat[]{\includegraphics[width=.48\linewidth]{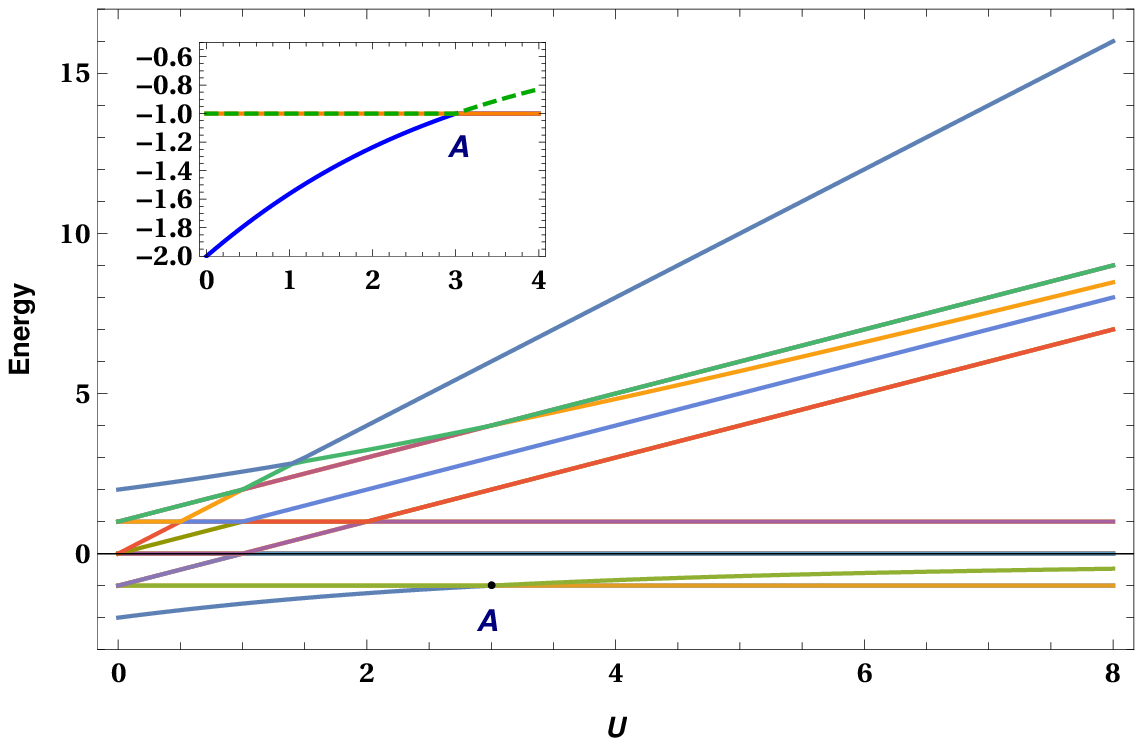} \label{fig:img1a}}\hfill
\subfloat[]{\includegraphics[width=.48\linewidth]{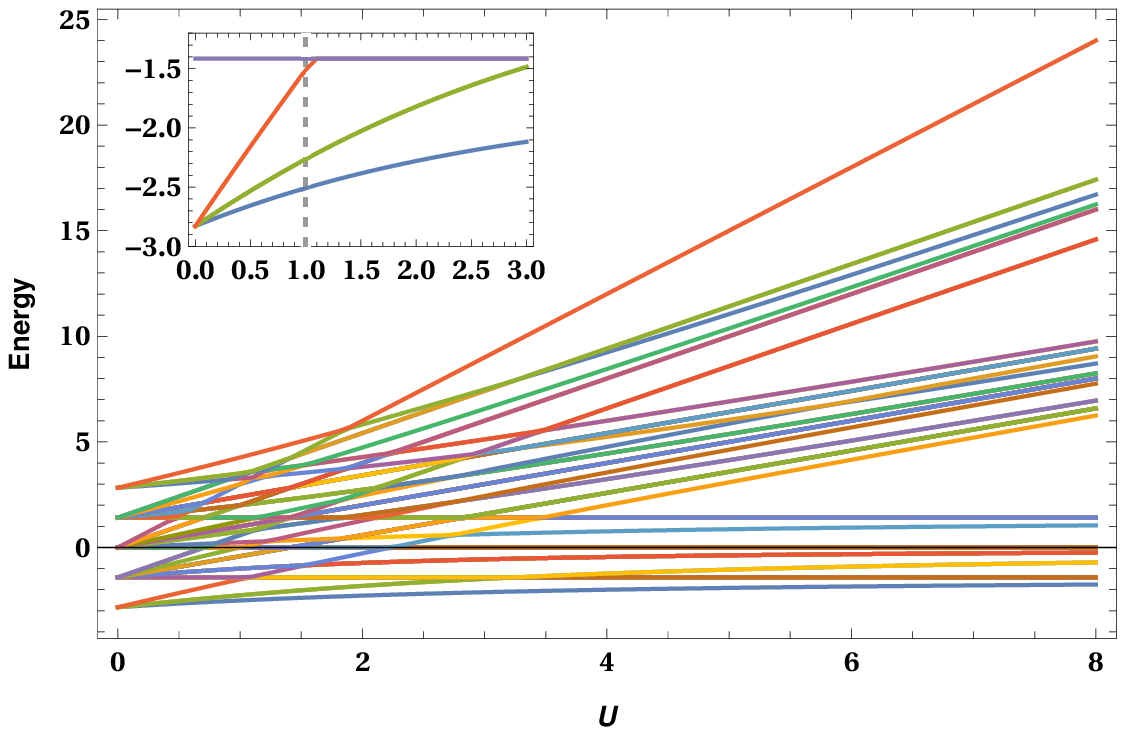}\label{fig:img1b}}
\caption{ \it The energy spectrum as a function of the Coulomb interaction $U$ for $N=2$ (a) and $N=3$ (b). The small graph on the top left shows the energy levels of the ground state (blue line) and the three lower exited states for $U$ between 0 and 3.  }
\label{fig:img1}
\end{figure}

In this section we present our results and their interpretation as pertaining to the one dimensional Fermi-Hubbard model with open boundary conditions. The behavior of different quantifiers introduced so far will be studied with respect to the different parameters of the system. Our interpretation of these results is based, among other things, on the behavior of the energy spectrum (see Figure  \ref{fig:img1}).

\subsection{Effect of Coulomb interaction and temperature on quantum correlations}

 We examine pair entanglement, pair coherence as well as pair mutual information under the effect of temperature ($k_{B}T/t$) and Coulomb interaction ($U= u/t$). We choose the dimensionless quantities $u/t$ and $k_{B}T/t$ as the two main variables to reduce the number of parameters of the model. Two main information will be especially of direct relation to the behavior of the correlation. On the one hand information about the ground state and its degeneracy and on the other hand the energy gap between the energy levels of the states.

From the spectrum corresponding to the 16 and the 64 eigenvalues displayed respectively in Figures (\ref{fig:img1a})  and (\ref{fig:img1b}), one can extract many information. The first observation of the spectrum shows that for small values of the Coulomb interaction, the energy levels overlap and the spectrum is narrow so the energy levels are close to each other. On the other end, with the increase of $U$, majority of the excited energy levels are moved up in energy such that, for strong coupling, the spectrum becomes broad and actually splits into several, separated energy bands. It is also interesting to notice that the number of these energy bands in the spectrum is equal to $N+1$ (the size of the chain plus one) with the highest one being comprised of a single energy level.  The lowest energy band is comprised of energy levels that are varying slowly, in contrast to the higher energy bands. It is also worthwhile noting that this behavior is related to the boundary conditions imposed on our system.

\subsubsection{Ground state and degeneracy}
\label{gsd}

The 1D Hubbard Hamiltonian (\ref{ham}) depends on the real parameter defined by the coupling constant $u$ or equivalently the dimensionless coupling constant $U$ defined earlier. Basing on this parameter we can make a distinction between two types of unusual degeneracies \cite{types}. The first one, depends on the coupling constant, which means that at some particular values of the coupling $U$ the degeneracy appears inducing thus a crossing of energy levels. This kind of degeneracy is clearly displayed in the energy spectrum when $U$ takes small values (Figure \ref{fig:img1}) where several intersection points connecting multiple energy levels (in the ground state and the excited states) appear at specific values of $U$. For instance, as displayed in Figure (\ref{fig:img1a}) for the simplest case of $N=2$, the ground state, corresponding to the smallest eigenvalue in the spectrum, is non-degenerate for $U <3$ but especially at $U=3$ the ground state becomes threefold degenerate \footnote{the crossing at point $A$ actually involves three energy levels as the first excited state is twofold degenerate before the point $A$ and becomes non-degenerate after the point $A$} while at $U>3$ the ground state turns into twofold degenerate.

The second type of degeneracy, can be called, the permanent one where the energy levels stay degenerate independently of the coupling constant $U$. This is highlighted in the spectrum when $U$ takes high values, where one can notice the disappearance of the crossings of the energy levels. The degeneracy of the ground state and the excited states then stabilizes and remains invariant for all values of $U$.

Generally, the high degree of degeneracy reflects the rich symmetry structure of the 1D Hubbard model\footnote{In fact, One can easily prove that by considering a simple example of an eigenstate $\ket{\Psi_{n}}$ of $H$ with energy $E_{n}$ ($H\ket{\Psi_{n}} = E_{n}\ket{\Psi_{n}}$). Knowing that $ S $ is a symmetry of the model (or equivalently $[H,S]=0$), then $S\ket{\Psi_{n}}$ is also an eigenstate of $H$ with the same energy $E_{n}$.

Therefore, due to the symmetry $S$, a degeneracy is induced in the energy level $E_{n}$ with two eigenstates $\ket{\Psi_{n}}$ and $S\ket{\Psi_{n}}$ ($\ket{\Psi_{n}} \neq S\ket{\Psi_{n}}$).}. Since the origin of degeneracy can be traced back to the symmetry of the system, one can thus define the $U$-dependent symmetries, that are comprised of Abelian symmetries as well as the Yangian non-Abelian symmetries, and the $U$-independent symmetries that are grouped into three classes, as specified by Heilmann and Lieb, a spatial symmetry associated to the lattice (such as mirror reflection symmetry for open boundary conditions ), spin symmetry and particle-hole symmetry  \cite{bookH}.

It is well known that temperature measures the degree of agitation associated to the particles constituting the matter. When one increases temperature the system acquires a thermal energy proportional to $k_{B}T$ ($k_{B}$ being the Boltzmann constant) which turns into kinetic energy (lattice vibrations) that is enough to transmit electrons from the lower energy level (ground state) to the higher energy levels (excited states).
At equilibrium, the thermal state describing the system at temperature $T$ is given by the density matrix
\begin{equation}
    \rho =\frac{e^{-\beta H}}{\Tr(e^{-\beta H})} ,
 \label{dens}    
\end{equation}
with $\beta= 1/k_{B} T$  and $H$ is the Hamiltonian defined in Eq.~\eqref{ham}.

If we consider discrete energy levels $E_{0}< E_{1}<E_{2}...$ of degeneracies $g_{0},g_{1},g_{2}...$ respectively, the density operator \eqref{dens} can be expressed in the energy eigenbasis as follows
\begin{equation}
    \rho=\sum_{i} P_{i} \ket{\psi_{i}}\bra{\psi_{i}} ,
\end{equation}
for which $P_{i}= \frac{e^{-\beta E_{i}}}{Z} $, $Z=\sum_{i} g_{i} e^{-\beta E_{i}}$ and $\sum_{i} P_{i} =1$.  

At zero temperature the state ~\eqref{dens} describing the system is exactly the ground state. In fact, for a given $i\ge 0$ the probability that the system occupies a level $E_{i}$ is 
\begin{equation}
  \begin{aligned}
     P_{i}  &= \frac{e^{-\beta E_{i}}}{Z} = \frac{e^{-\beta E_{i}}}{\sum_{j} g_{j} e^{-\beta E_{j}}}  =  \frac{e^{-\beta (E_{i}-E_{0})}}{\sum_{j}g_{j} e^{-\beta (E_{j}-E_{0})}} \\ & =\frac{e^{-\beta (E_{i}-E_{0})}}{g_{0}+\sum_{j>0} g_{j} e^{-\beta (E_{j}-E_{0})}}.
  \end{aligned}    
\end{equation}
When $ T \rightarrow 0$  (equivalently  $\beta \rightarrow +\infty$),
\begin{equation}
    \lim\limits_{\beta \rightarrow +\infty} P_{i}= \frac{\lim\limits_{\beta \rightarrow +\infty} e^{-\beta (E_{i}-E_{0})}}{g_{0}+ \lim\limits_{\beta \rightarrow +\infty} \sum_{j>0} g_{j} e^{-\beta (E_{j}-E_{0})}}= \frac{\delta_{i0}}{g_{0}}.
 \label{pi}
\end{equation}
In other words Eq.~\eqref{pi} asserts that at zero temperature, all the Boltzmann weights $e^{-\beta E_{i}}$ of the density $\rho$ for which $i\neq 0$ vanish and only the weight corresponding to the ground level $E_{0}$ persists and the state becomes $\rho= \frac{1}{g_{0}} \sum_{i=1}^{g_{0}}\ket{\psi_{i}}\bra{\psi_{i}}$. If $g_{0}=1$ the ground state is pure otherwise it is mixed.

If we increase temperature the systems may move from the ground level to the excited levels, which means that more weights $e^{-\beta E_{i}}$ with $i>0$ will be added to the density matrix $\rho$, which also implies that the system will get more thermal mixture and generally this is accompanied with a decay of the correlations (entanglement, coherence,...).

\subsubsection{The state energy difference  $\Delta E$}

An essential factor that influences the transition to the excited states is the energy gap between the relevant states. In order to take this into account, we define the ratio, also known as the Boltzmann factor,
\begin{equation}
  \frac{ P_{u}}{ P_{l}}=  \frac{ N_{u}}{ N_{l}}=  \frac{ g_{u}}{ g_{l}}e^{-\beta \Delta E },
\end{equation}
where $N_{l}$ and $N_{u}$ are respectively the number of particles occupying the lower level and the upper level. $\Delta E= E_{u}-E_{l}$ is the state energy difference or the gap between the energy levels. If the system admits only these two levels its state is described by the following density matrix
\begin{equation}
    \rho=\frac{e^{\beta \Delta E}}{1+e^{\beta \Delta E}} \ket{\psi_{l}}\bra{\psi_{l}}+ \frac{e^{-\beta \Delta E}}{1+e^{-\beta \Delta E}}\ket{\psi_{u}}\bra{\psi_{u}}.
\end{equation}
When $\Delta E >> 1$ the upper level will always have a lower probability of being occupied, but instead it has a relatively higher probability to be occupied when $\Delta E << 1$. In the latter case, the density matrix is represented as a mixture of the two Boltzmann weights $e^{-\beta E_{l}}$ and $e^{-\beta E_{u}}$, which consequently decreases the entanglement in the system.

\subsection{Correlations at zero temperature}
\label{czt}

\begin{figure}[ht]\centering
\subfloat[]{\includegraphics[width=.45\linewidth]{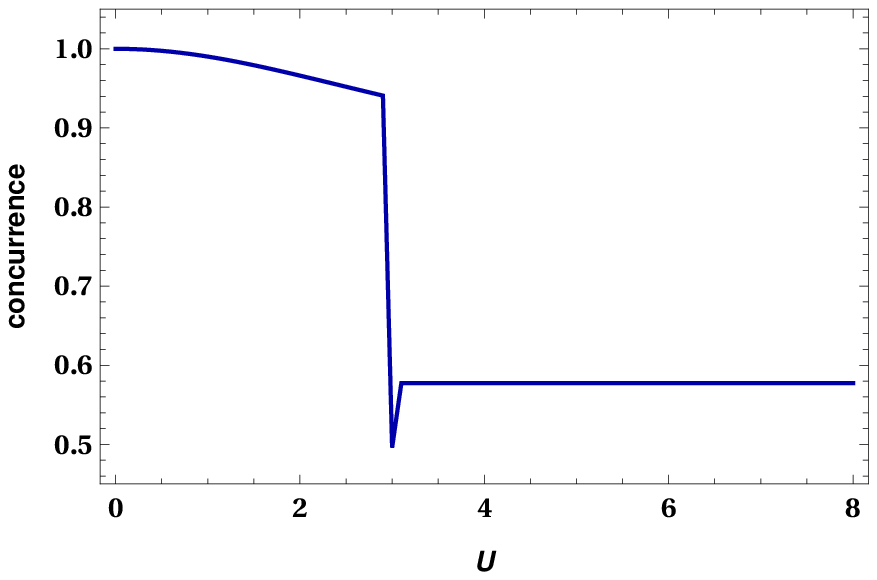}\label{fig:img2a}}\hfill
\subfloat[]{\includegraphics[width=.45\linewidth]{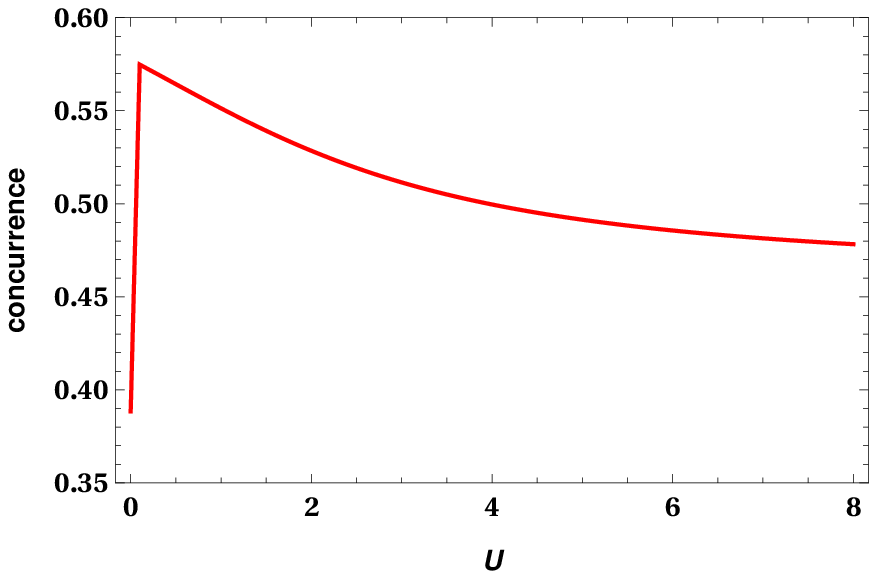}\label{fig:img2b}}
\caption{ \it The lower bound of concurrence corresponding to the ground state as a function of $U$ for  $N=2$ (a)  and for the pair $\rho_{1,2} = \rho_{2,3}$ for  $N=3$ (b). }
\label{fig:img2}
\end{figure}

As argued in the previous subsection, the correlations at zero temperature are exactly the correlations of the ground state. Since the degeneracy of this latter depends on the coupling constant $U$ and the size $N$ of the chain, the entanglement, coherence and the total correlations of the ground state will be strongly affected too by these same parameters. For the entanglement, this can be clearly seen in Figure \ref{fig:img2} where we plot the lower bound of concurrence (\textit{LBC}) for $N=2$ in (\ref{fig:img2a}) and for $N=3$ in (\ref{fig:img2b}).

For $N=2$, Figure (\ref{fig:img2a}), we see that the $LBC$ is gradually decreasing as long as $U<3$, then there is a sudden jump in its behavior exactly at $U=3$ and it stabilizes afterward. This behavior can be interpreted in conjunction with the behavior observed in Figure (\ref{fig:img1a}). This latter allows to observe that the ground state is non-degenerate when $U<3$, which is the interval in which the entanglement diminishes with $U$ in Figure (\ref{fig:img2a}). At $U=3$, where the ground state is threefold degenerate and thus represented by the mixture $\rho=\sum_{i=1}^{3} P_{i}\ket{\psi_{i}}\bra{\psi_{i}}$ the entanglement (more precisely the $LBC$) undergoes an abrupt decrease that goes (from almost maximal entanglement) to $0.50$. For $U>3$ the ground state is twofold degenerate as we discussed in subsection \ref{gsd}, then the state has less mixture ($\sum_{i=1}^{2} P_{i}\ket{\psi_{i}}\bra{\psi_{i}}$) compared to the previous case so the entanglement increases again abruptly to about 0.58. With the increase of $U$, the strong repulsion interaction allows electrons to move away from each other (thus sites with double occupancy $\ket{{\color{red}\uparrow} {\color{blue}\downarrow}}$ are excluded) without being able to tunnel between the quantum dots. The confinement state reached is given by $\rho= \frac{1}{2}\ket{\phi_{1}}\bra{\phi_{1}}+\frac{1}{2}\ket{\phi_{2}}\bra{\phi_{2}}$, with $\ket{\phi_{1}} = \frac{1}{\sqrt{2}}(\ket{0{\color{red}\uparrow}}+\ket{{\color{red}\uparrow} 0})$ and $ \ket{\phi_{2}} = \frac{1}{\sqrt{2}}(\ket{0{\color{blue}\downarrow}}+\ket{{\color{blue}\downarrow} 0})$. This confinement state reached is independent of $U$, for $U>3$ (as the degeneracy becomes also independent in that interval) and it is for this reason that entanglement ($LBC$) is constant as depicted in Figure (\ref{fig:img2a}) .

For $N=3$, (Figure (\ref{fig:img2b})), the ground state is four times degenerate at $U=0$ \footnote{generally, at $U=0$ the ground state is non-degenerate for $N$ even but degenerate for $N$ odd.} but for $U>0$ the state becomes non-degenerate, therefore the mixture disappears and this is the reason why entanglement of the pair $\rho_{1,2} $ or\footnote{because of the mirror reflection symmetry, the two bipartite states $\rho_{1,2}$ and $\rho_{2,3}$ are equal.} $\rho_{2,3} $ has an abrupt increase from 0.38 to 0.55. By the same reasoning mentioned above, the confinement state in the strong coupling region tends to $\ket{\phi}= \frac{1}{\sqrt{8}}(\ket{0{\color{blue}\downarrow}{\color{red}\uparrow}}-\ket{0{\color{red}\uparrow}{\color{blue}\downarrow}}+\ket{{\color{blue}\downarrow}{\color{red}\uparrow}0}-\ket{{\color{red}\uparrow}{\color{blue}\downarrow}0})+\frac{1}{2}(\ket{{\color{red}\uparrow}0{\color{blue}\downarrow}}-\ket{{\color{blue}\downarrow}0{\color{red}\uparrow}})$ which makes entanglement asymptotically stabilize for high coupling values.

This reasoning allows to explain also the behavior of the quantum coherence and the mutual information, of which we chose not to include the plots \textit{vs} the coupling constant $U$ as their behavior is similar to that of the $LBC$. Instead, we show in the next subsection their behavior as a function of temperature for different values of $U$ in Figure \ref{fig:img4}. As discussed earlier, at $T=0$ this is equivalent to studying the ground state  behavior. We notice that mutual information is quantitatively higher than coherence and entanglement as expected because the mutual information encompasses all the classical and quantum correlations of the system. Also, It is clear that the coherence in the system steams only from the quantum correlations shared between the two sites, because as mentioned in section \ref{mesure} the one site density matrix Eq.~\eqref{gho} is diagonal which means that there is no local coherence.

\subsection{Correlations at finite temperature}

\begin{figure}[ht]\centering
\subfloat[]{\includegraphics[width=.48\linewidth]{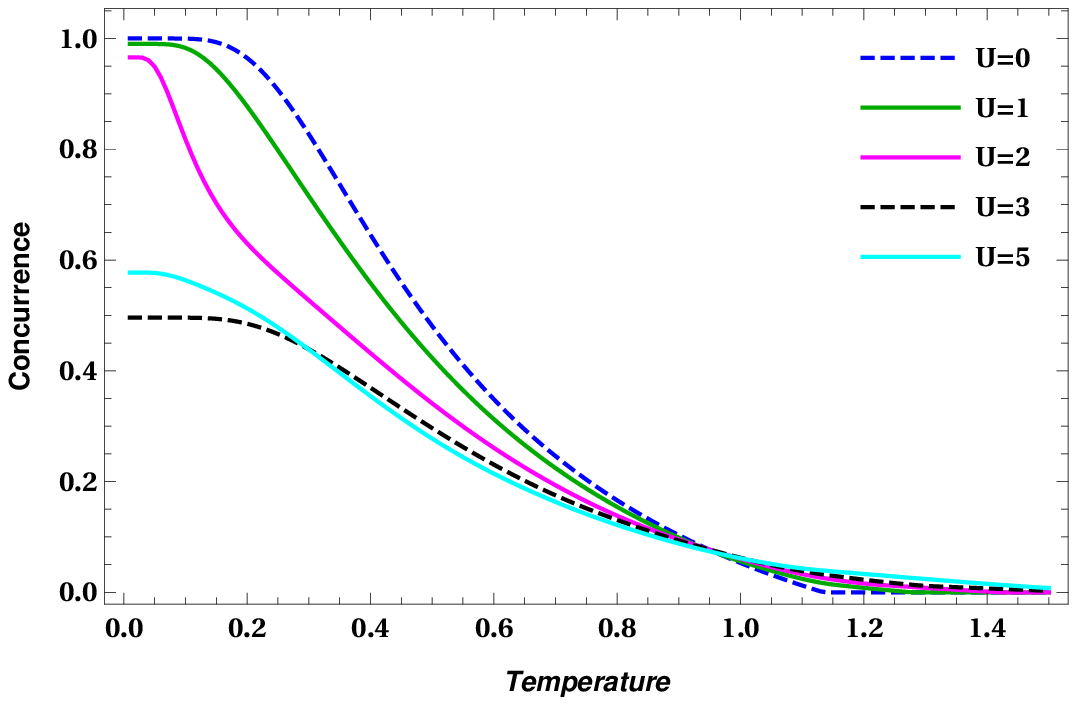} \label{fig:img3a}}\hfill
\subfloat[]{\includegraphics[width=.48\linewidth]{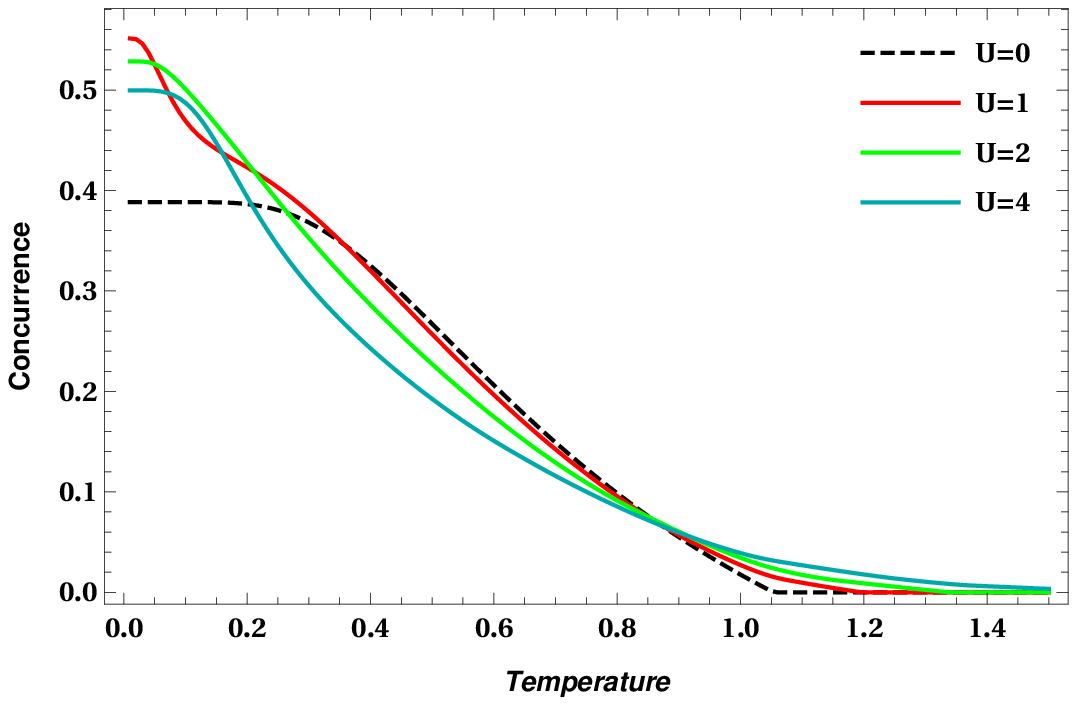}\label{fig:img3b}}\par 
\subfloat[]{\includegraphics[width=.48\linewidth]{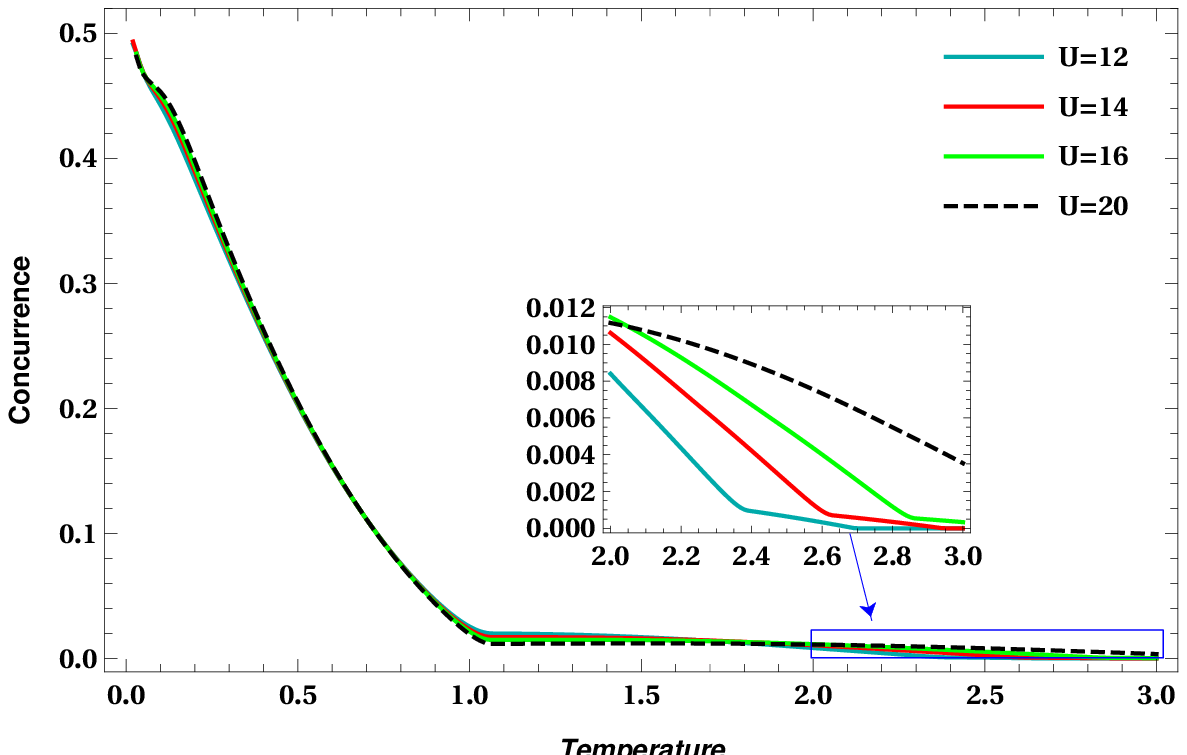} \label{fig:img3c}}
\caption{Lower bound of concurrence in the state $\rho_{1,2}$ as a function of temperature ($k_{B}T$) for different values of the Coulomb interaction $U$ and with size $N=2$ (a),  $N=3$  (b),  $N=4$ (c).  }
\label{fig:img3}
\end{figure}

At finite temperature the behavior of correlations depends essentially on the energy difference $\Delta E$ between the ground state and the nearest excited states. The spectrum of $N=2$, (Figure (\ref{fig:img1a})), shows that $\Delta E$ between the ground state and the first excited state decreases with $U$ as long as $U<3$. This means that the first excited state has more probability to be populated as $U$ increases. Therefore the mixture increases too and this leads to the entanglement's rapid decrease under the effect of small temperatures when $U$ increases. At $U=3$ the difference $\Delta E$ is higher between the ground state (degenerate at this value) and the nearest excited state so in this case the entanglement decreases very slowly and a constant rate persists for small temperatures.

This effect is also observed for $N=3$ with the pair $\rho_{1,2}$. Indeed, the spectrum (Figure (\ref{fig:img1b})) shows that $\Delta E$, between the ground state and the nearest excited state, increases with $U$  (for relatively small values of $U>0$), thus the exited state has less chance to be populated and to be in a thermal mixture. This explains the slow decay of entanglement with the increase of $U$ at small temperatures (Figure (\ref{fig:img3b})). At $U=0$, $\Delta E$ between the ground state and the first excited state is higher (see Figure (\ref{fig:img1b}) and the zoomed plot within), consequently entanglement stays constant at small temperatures and decays slowly with the increase of temperature.

To fill more excited states we have to increase temperature and here we notice two special behaviors. 
The first is observed in Figures (\ref{fig:img3a}) and (\ref{fig:img3b}), where a change in the rate of decrease of entanglement with temperature is obviously remarked for some values of $U$\footnote{namely, for $U=2$ in Figure (\ref{fig:img3a}) and for $U=1$ in Figure (\ref{fig:img3b}).}. This can be explained following the same reasoning mentioned before, \textit{i.e.} that this is due to the widening of $\Delta E$ between lower neighboring excited states. As a matter of fact, if we take the example $U=1$ for $N=3$, the spectrum in Figure (\ref{fig:img1b}) and the subplot within, show that the quantity $\Delta E$ between the first and the second excited state is larger compared to $\Delta E$ between the ground state and the first excited state. In this regard the probability that the system occupies the second excited state is very low, therefore entanglement starts to decay slowly and this explains the change in the rate of decrease of entanglement in Figure (\ref{fig:img3b}) for $U=1$. The same thing applies to $U=2$ for $N=2$ in Figures (\ref{fig:img1a}) and (\ref{fig:img3a}). In contrast to this, for the other values of $U$ and for mid temperatures there is no change in the rate of decrease of entanglement with temperature (Figure \ref{fig:img3}) as the energy difference $\Delta E$ between the lower states is comparable (Figure \ref{fig:img1}).

The second behavior is observed for the strong coupling $U$ in Figure (\ref{fig:img3c}) where the curves converge toward each other. As observed in subsection \ref{czt}, the ground state entanglement stabilizes for large values of $U$ (Figure \ref{fig:img2}) as the system reaches a confinement state. So the increase in $U$ has little effect on the entanglement and this carries out also as the temperature increases which is seen by the very small discrepancy in the plots for different values of $U$ in Figure (\ref{fig:img3c}). This is confirmed by noticing, from Figure \ref{fig:img1}, that the energy levels in the first band are very close to each other and that the quantity $\Delta E$ between the levels is approximately the same. This happens at the first band as the thermal energy supplied to the system is not yet enough to move the system up to the upper bands.

\begin{figure}[ht]\centering
\subfloat[N=2]{\includegraphics[width=.47\linewidth]{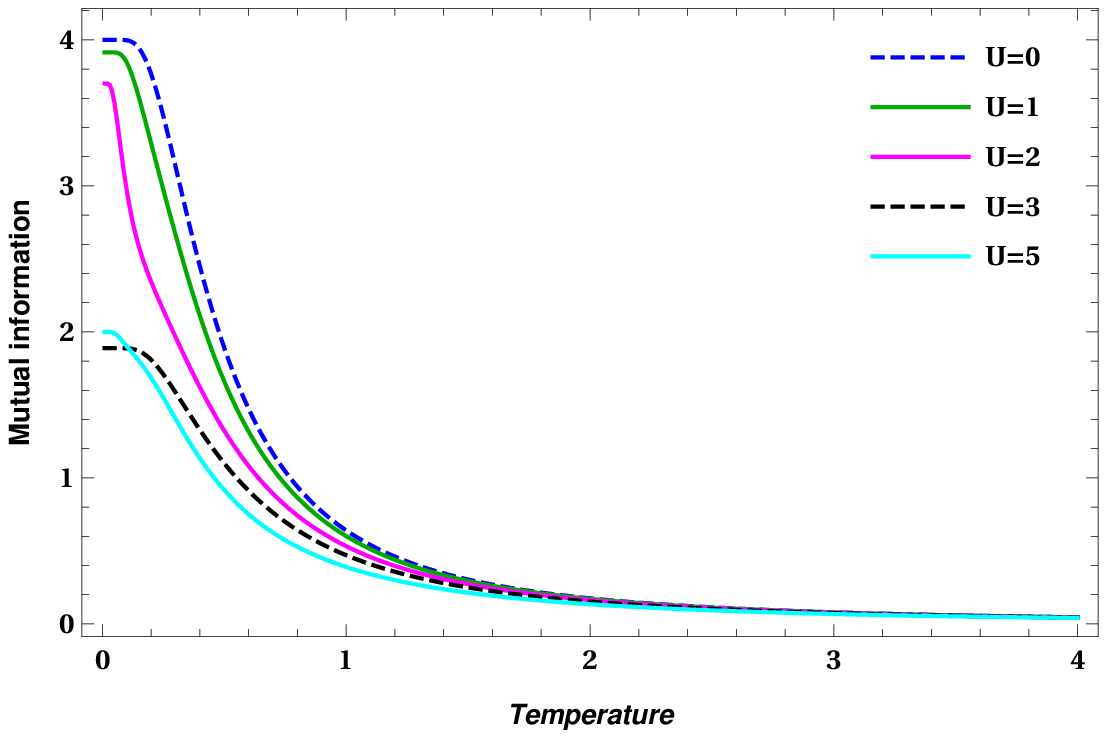}\label{fig:img4a}}\hfill
\subfloat[N=3]{\includegraphics[width=.48\linewidth]{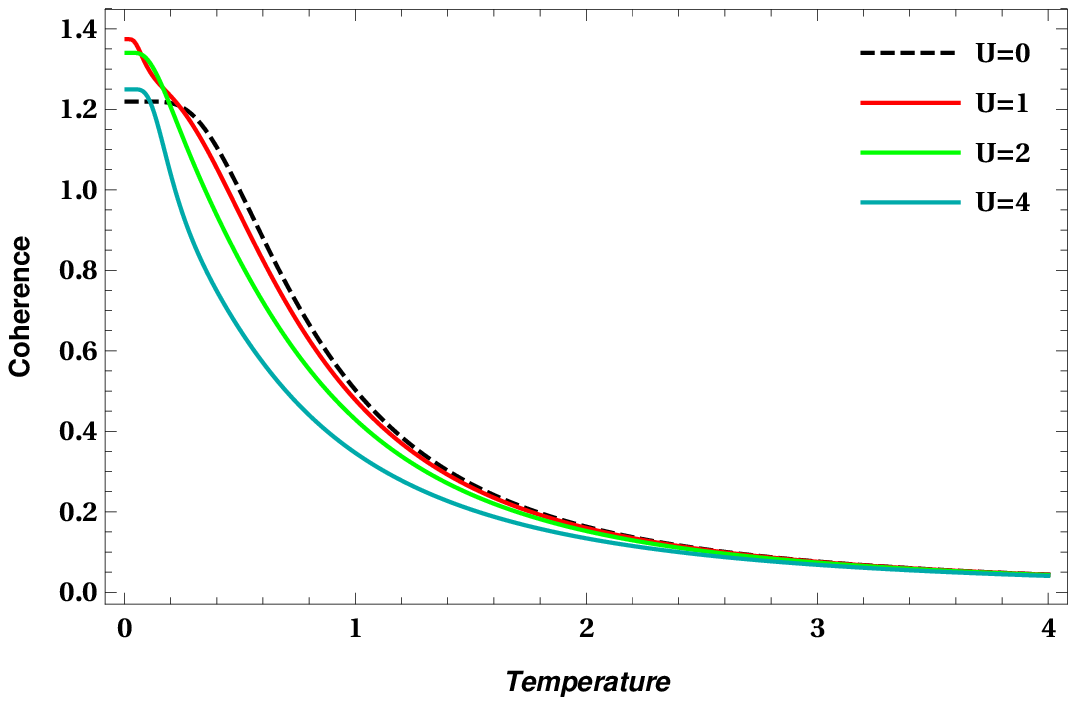}\label{fig:img4b}}
\caption{ \it Total correlations (a) and Coherence (b) as a function of temperature for different values of $U$. }
\label{fig:img4}
\end{figure}

For completeness, we show the behavior of the  mutual information and the quantum coherence at finite temperature in Figures (\ref{fig:img4a}) and (\ref{fig:img4b}) respectively. 

Generally the behavior is similar to that of entanglement as is easily observed by comparing with Figure \ref{fig:img3}. The same interpretation in terms of the ground state's degeneracy for $T=0$ (see discussion at the end of subsection \ref{czt}) and in terms of the energy difference $\Delta E$ for finite temperature still applies here also.

\subsection{Correlations at high temperature}

Generally at high temperature the system has enough thermal energy to occupy the higher exited states (also those in the higher bands). The entanglement (as well as the other correlations like quantum coherence and mutual information) thus vanishes due to the fact that the state turns into a mixture of all the possible Boltzmann wights (thermal relaxation effect). 

The effect of the Coulomb interaction, as seen from Figure \ref{fig:img3}, is that for a given pair $\rho_{i,j}$ at high temperature, the entanglement survives longer when increasing the value of $U$; only very high temperatures are able to destroy entanglement when $U$ takes very large values. As before, the explanation of this is obtained by going back to the energy spectrum in Figure \ref{fig:img1}. Indeed, the gap separating the bands in the spectrum increases dramatically with $U$ which makes it difficult for the system to transition to higher energy levels and bands. Therefore more thermal energy is needed to overcome this and obtain a mixture of all the possible Boltzmann wights resulting eventually in the vanishing of entanglement. In other words, the high Coulomb interaction between electrons makes the state strongly entangled, and hence at high temperature the thermal fluctuations are in competition with these strong entangled correlations.

In addition to the behavior and related interpretation given above, that still applies to quantum coherence and mutual information, Figure \ref{fig:img4} shows that these correlations are more robust against temperature compared to entanglement and higher temperatures (for all values of $U$) are needed to destroy these correlations. This is a recurring remark when comparing the robustness of entanglement \textit{vs} that of more general correlations (coherence, quantum discord...).

\section{Conclusion} 

Quantum dot systems usually are treated in quantum information as qubit systems whenever the study of entanglement concerns the mixed state case  \cite{ Urbaniak_2013, hich, coupledQD}. Notably, description of quantum dots in terms of the Hubbard model as qubits is valid only in some specific cases requiring more restrictive constraints. In this paper we adopted the less restrictive point of view, but yet more accurate, in describing a system of quantum dots as a 1D Fermi-Hubbard chain \cite{approximationqdots}. In this approach a quantum dot is described by a quadrit (object described by a four dimensional Hilbert space) instead of a qubit.

The absence of properly defined measures of mixed state entanglement for qudits (objects with higher dimensional Hilbert spaces) and for quadrits in particular forces us to adopt a novel approach in tackling questions related to quantum dots and the quantum correlations present in these systems. Namely, we opted to study the behavior of pairwise entanglement by calculating the lower bound of concurrence. The general conclusions were then confirmed by comparing the behavior of more general types of correlations.

In summary, The energy spectrum of the system provided a proper explanation and interpretation of the different results and behaviors observed. As a matter of fact, the  entanglement of the ground state for $N=2$ and $N=3$ was studied and the influence of regions where we have the $U$-dependent and  $U$-independent degeneracies was established.

Furthermore we have studied the influence of the state energy difference $\Delta E$ on the decay of correlations. This study allows to explain the behavior of entanglement at finite and high temperature and establishing that at finite temperature, entanglement as well as quantum coherence and total correlations decay rapidly when $\Delta E$ is very small but when $\Delta E$ is high the correlations decay very slowly with temperature. In addition, we have demonstrated that at high temperature the stronger the Coulomb interaction (compared to the tunneling of electrons), the stronger the entanglement, 
which gives rise to a long survival of entanglement due to the broad energy difference $\Delta E$ separating the bands in the region when $U$ is high.

\section*{Acknowledgements}

S. Abaach acknowledges support from the National Center for Scientific and Technical Research (CNRST). 

The authors would like to thank Andreas Buchleitner and Edoardo Carnio for fruitful discussions.

\bibliography{sanaa}

\end{document}